\documentclass[twocolumn,prb,10pt,showpacs]{revtex4}
\usepackage[utf8]{inputenc}
\usepackage{graphicx}
\usepackage{amsmath}
\usepackage{amssymb}
\usepackage{bm}
\usepackage{hyperref}
\usepackage{color}
\usepackage{braket}
\usepackage{dsfont}

\begin{document}

\title{Tomography of  Majorana Fermions with STM Tips}
\author{Denis Chevallier and Jelena Klinovaja}
\affiliation{Department of Physics, University of Basel, Klingelbergstrasse 82, CH-4056 Basel, Switzerland}

\date{\today}
\begin{abstract}
We investigate numerically the possibility to detect the spatial profile of Majorana fermions (MFs) by using STM tips that are made of either normal or superconducting material. In both cases, we are able to resolve the localization length and the oscillation period of the MF wavefunction.
We show that the tunneling between the substrate and the tip, necessary to get the information on the wave function oscillations, has to be weaker in the case of a superconducting probe. In the strong tunneling regime, the differential conductance saturates  making it more difficult to observe the exponential decay of MFs.
The temperature broadening of the profile is strongly suppressed in case of the superconducting tip resulting, generally, in better resolution. 
\end{abstract}


\maketitle

\section{Introduction}

Majorana fermions (MFs) have been intensively studied in different condensed matter systems during the last decade \cite{Fu,MF_Sato,MF_Sarma,MF_Oreg,alicea_majoranas_2010,potter_majoranas_2011,
Klinovaja_CNT,Pascal,Bena_MF,Fra,Rotating_field,Ali,RKKY_Basel,RKKY_Simon,RKKY_Franz,Pientka,Ojanen,Carlos}.
In partial, these states are interesting due to their exotic properties such as non-Abelian statistics, which open the perspective of using them for quantum computing \cite{Alicea,Flensberg,Beenakker,Kitaev,Adrian,Fabio}. Experimental evidences of such states have been reported in semiconducting nanowires with strong Rashba spin-orbit interaction (SOI) brought in the proximity to an $s$-wave superconductor (SC) \cite{Mourik,Marcus,Heiblum,Us} and in magnetic atomic chains on SC substrates \cite{Franke,Yazdani,Meyer}. In contrast to transport experiments \cite{Mourik,Marcus,Heiblum,Us}, where one can only confirm the presence of zero-energy states but not its localized character in real space, the more recent scanning tunneling microscopy (STM) experiments accessed the MF wavefunction showing that the observed zero-energy states are localized at the chain ends \cite{Franke,Yazdani,Meyer}. However, a systematic numerical study of the full tomography of the MF wavefunction using STM techniques is still missing. 
In this paper we focus on the tomography of the MF wave function by modeling  STM tips made either of normal or superconducting materials and compare both approaches. 

We find that the differential conductance is always position-dependent. Generally, in the weak coupling regime, maximums (minimums) in the conductance correspond to maximums (minimums) in the local density-of-states (LDOS), which allows us to access the MF wavefunction properties such as the localization length and the spatial oscillation period. In the strong coupling regime, the differential conductance saturates and never exceeds the quantized values $G_{sc}=(4-\pi)2e^2/h$ [$G_{n}=2e^2/h$] for the superconducting (normal) tip at low temperatures \cite{Glazman,Law,Anton}.  If temperature is high, the thermal broadening becomes important, and the maximum conductance is much lower than $G_{sc}$ or $G_{n}$ as well as in the case when the coupling between the STM tip and the substrate is weak.  We also find two important advantages in using a superconducting over a normal tip. First, the tunneling rate necessary to get the information on the wave function has to be smaller than for a normal STM tip. Second, the temperature broadening is strongly suppressed. We finally discuss several effects such as the resolution of the tip compared to the period of the oscillations, and also various model regimes allowing us to control and detect the properties of MF wave functions in the most optimal regime.

Our calculations are based on the non-equilibrium Green function technique involving the Keldysh formalism. We calculate the bare Green function of our substrate which is dressed by the self energy of the tip via a non-perturbative tunneling term \cite{Zazunov, Chevallier}.  Next we get the full Green function of the system which allows us to derive a general formula for the current and the higher cumulants. Importantly, this model is quite general and could be used to take into account, for instance, the Coulomb interaction in the nanowire \cite{Chevallier,Chevallier2}.

\begin{figure}[t]
\includegraphics[width=8cm]{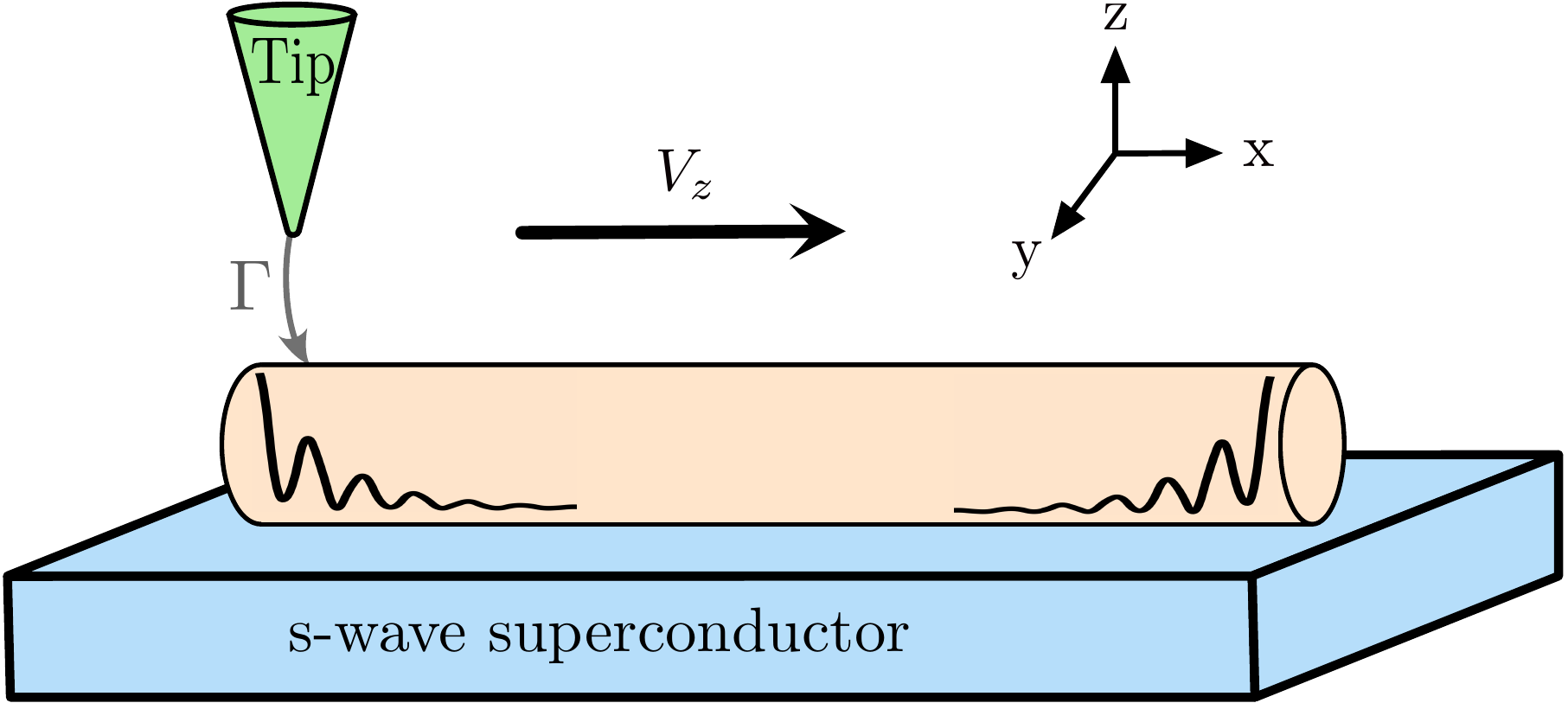}
\caption{Schematics of the setup: One-dimensional Rashba nanowire is aligned along the $x$-axis and placed on top of an $s$-wave superconductor. An external magnetic field $V_z$ is applied in the $x$ direction. A current via the STM tip, which is weakly coupled to the substrate with the tunneling strength $\Gamma$, allows one to confirm the presence of MF states and, in addition, to get information about their spatial profile.}
\label{fig:Fig1}
\end{figure}

The paper is organized as follows. In Sec. \ref{Model} we present our model. In Sec. \ref{Normal} and \ref{Superconducting}, we study the detection using a STM tip made of, respectively, normal metal and superconducting metal. Finally, we discuss some additional aspects of the detection in Sec. \ref{Discussions}.

\section{Model}\label{Model}

\begin{figure}[t]
\includegraphics[width=8cm]{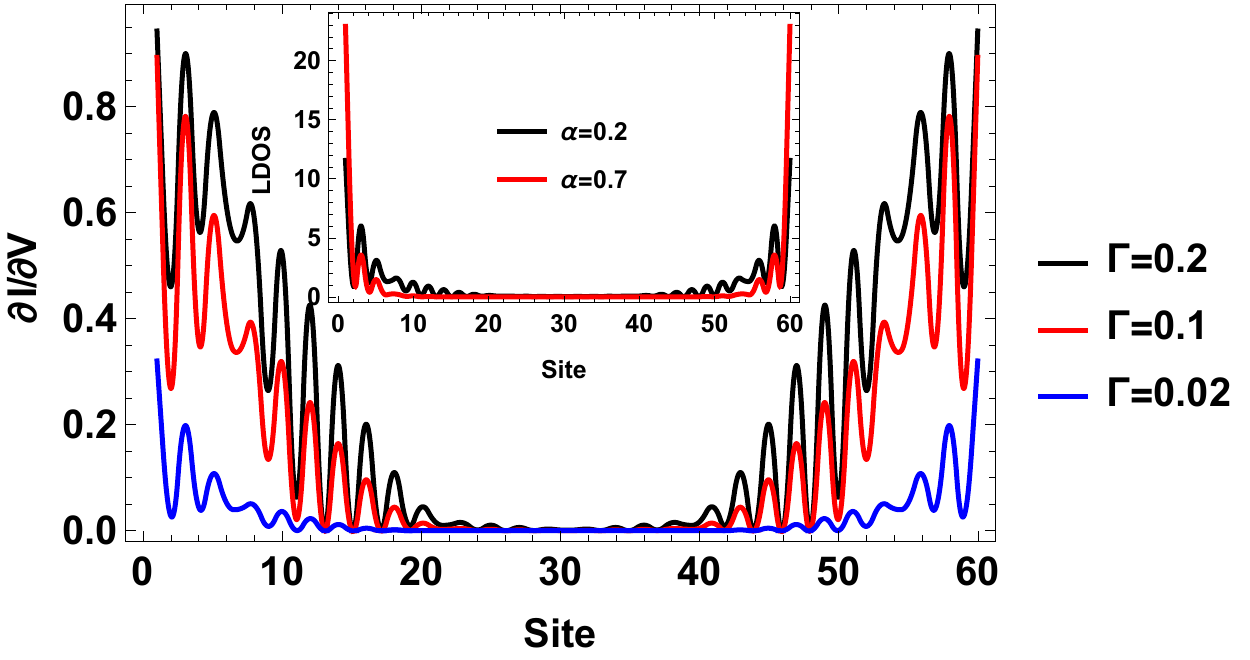}
\caption{Differential conductance (normalised to $G_n$ in what follows) obtained with the normal tip at zero bias as a function of position for three values of the transmission $\Gamma$ at $1/k_{B}T=200$ and when the substrate is a $N=60$ sites length in the following configuration : $\alpha=0.2$, $\mu=2$, $V_z=2$, $\Delta=1$ (in units of $\hat{t}=1$). The spatial conductance maps allow one to resolve both MF localization lengths and the oscillation period of MF wavefunctions. For comparison, in the inset, we show the zero energy LDOS as a function of the position for two values of $\alpha$. The stronger the SOI is, the more localized are the MFs.}
\label{fig:Fig2}
\end{figure}

Our setup consists of two parts, namely, the substrate hosting  MFs and the STM tip (either in normal or superconducting state) that allows one to probe the LDOS of the substrate. As a substrate we consider a one-dimensional Rashba nanowire  aligned along the $x$-axis brought into contact with an $s$-wave superconductor in presence of an external magnetic field applied in the $x$ direction (see Fig. \ref{fig:Fig1}).  Numerically, we describe the nanowire in the tight-binding model framework. The  corresponding Bogoliubov-de-Gennes Hamiltonian is written in the Nambu basis as, 
\begin{align}\label{substrate_hamiltonian}
\tilde{H} = \sum^N_{j=1} \tilde{\psi}^\dagger_j &\left[-\mu\tau_z+\Delta \tau_x+V_z \sigma_x\right]\tilde{\psi}_j\notag\\
&+\sum^{N-1}_{j=1} \tilde{\psi}^\dagger_{j+1}\left[-\hat{t}-i\alpha \sigma_y\right]\tau_z\tilde{\psi}_j +H.c. ,
\end{align}
where
$\tilde{\psi}_j= (\psi^{\dagger}_{j,\uparrow}, \psi^{\dagger}_{j,\downarrow},  \psi_{j,\downarrow} , -\psi_{j,\uparrow})$, $N$ is the number of lattice sites, and the Pauli matrices $\sigma_i$ ($\tau_i$) act on spin (particle-hole) space. The operator $\psi^{\dagger}_{j,\sigma}$ creates a particle of spin $\sigma$ at site $j$. Here, $\mu$ is the chemical potential, $\hat{t}$ the hopping strength, $\Delta$ the $s$-wave superconducting pairing amplitude assumed to be induced by proximity effect\footnote{Alternatively, one can take into account the bulk $s$-wave superconductor directly in the self-energy and, as a result, in the nanowire Green function \cite{Sasha}. This will allow one to introduce the proximity-induced gap straightforwardly. For simplicity, in our simulations, we work with the effective superconducting Hamiltonian. Of course, if the coupling between the nanowire and the substrate is strong, the MF wavefunction will leak into the bulk superconductor, and, thus, the STM signal will be finite also over the substrate.}, $\alpha$ the strength of SOI, and $V_z$ is Zeeman energy. In the topological regime, such chains support zero-energy modes localized at the end of the nanowire \cite{MF_Sarma,MF_Oreg}. In our formalism we can confirm the presence of MFs by calculating the LDOS at zero energy at a given position $j$ along the nanowire (see inset of Fig. \ref{fig:Fig2} and the Appendix \ref{3Dplots})
\begin{equation}
\rho_j(\omega) = -\frac{1}{\pi}\sum_{\sigma=\uparrow,\downarrow}\textrm{Im}[\tilde{G}_{0R} (\omega)]_{jj,\sigma\sigma},
\end{equation}
where the Green function of the substrate alone is defined as $\tilde{G}^{-1}_{0R} (\omega)=\omega+i \delta-\tilde{H}$, with $\delta$ being an infinitesimal which allows us to invert this matrix. Generally, the localization length of a MF is inverse proportional to the gap in the spectrum and depends on system parameters \cite{Alicea,JK_MF}. Importantly, we choose the length of the nanowire such that two MFs do not overlap with each other to avoid any possible splitting \cite{Sasha, Ramon}.

The Hamiltonian for the superconducting STM tip is written as
\begin{equation}
\tilde{H}_{tip}=\sum_{k,\sigma} \xi_k \Psi^\dagger_{k,\sigma}\Psi_{k,\sigma}+\sum_{k }(\Delta_{tip} \Psi^\dagger_{k,\uparrow} \Psi^\dagger_{-k,\downarrow} +H.c.),
\end{equation} 
with $\xi_k  = k^2/2m-\mu_s$ and $\Psi_{k,\sigma}$ being the annihilation operator of an electron in the tip with spin $\sigma$ and momentum $k$.  The normal metal STM tip is obtained by setting $\Delta_{tip}=0$. The tunneling Hamiltonian between the tip and the nanowire is written as $\tilde{H}_{T}=\sum_{k}\tilde{\Psi}^\dagger_{k} \tilde{t}_j \tilde{\psi}_{j} e^{ikj}+H.c.$ where $\tilde{\Psi}^\dagger_{k} $ corresponds to the Nambu spinor composed of electron operators of the STM tip. In what follows, we consider that the tunneling occurs between the tip and the site $j$ of the nanowire for which $\tilde t_j >0$. The voltage difference between the tip and the substrate is included in the tunneling parameter via a Peierls transformation $\tilde{t}_j=\hat{t}_j\tau_z e^{i \tau_z V t}$ with $t$ denoting the time 
\footnote{We checked that applying a symmetric bias, namely $+V/2$ on the tip and $-V/2$ on the substrate does not affect the results for the differential conductance.}. Since the total Hamiltonian is quadratic in the tip degrees of freedom, we can integrate out these modes, such that the effect of the tip is taken into account in the self-energy $\tilde{\Sigma}_{R}(\omega)$ that dresses the bare Green's function of the nanowire $\tilde{G}^{-1}_{R}(\omega) = \tilde{G}^{-1}_{0R}(\omega)-\tilde{\Sigma}_R(\omega)$. The total retarded self-energy can be written in the space of sites as $[\tilde{\Sigma}_R(\omega)]_{ii}=\tilde{\Sigma}_{i,R}(\omega)$ where the components are non-zero only at the site $i=j$.
 The on-site retarded self-energy of the tip is given by $\tilde{\Sigma}_{j,R}(\omega)=\Gamma_j \tau_z \tilde{g}_R (\omega) \tau_z$, where $\tilde{g}_R (\omega)$ is its retarded Green function and $\Gamma_j=\pi \nu(0)|\hat{t}_j|^2$ the tunneling rate.  Using the total Green function and the self-energy of the tip, we calculate the current using the Keldysh formalism (see the Appendix \ref{Model_derivation} and \cite{Chevallier}),
\begin{equation}
I_{dc}=\frac{e}{2\hbar}\textrm{Tr}\{\tau_z \int^{+\infty}_{-\infty} \frac{d\omega}{2\pi} [\tilde{G}_R (\omega)\tilde{\Sigma}_K(\omega)+\tilde{G}_K(\omega)\tilde{\Sigma}_A(\omega)]\}.
\end{equation}
The corresponding differential conductance is computed 
as $G={\partial I_{dc}}/{\partial V}$. 

\begin{figure}[t]
\includegraphics[width=8cm]{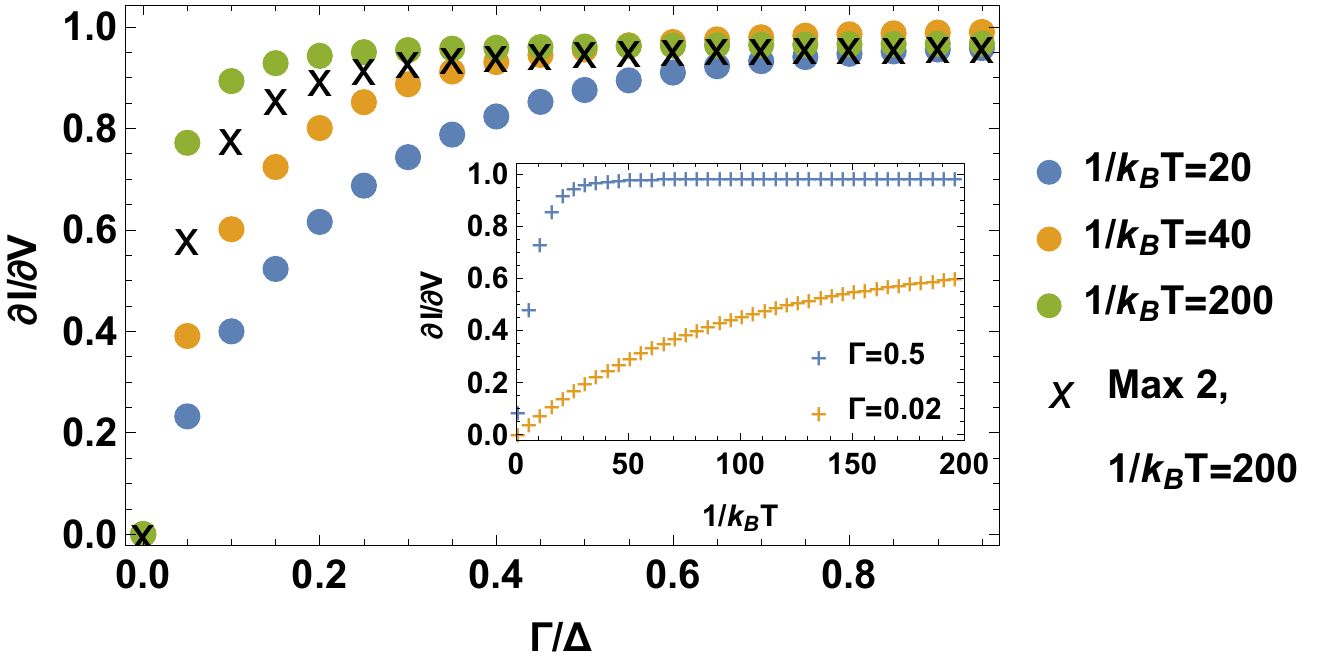}
\caption{Differential conductance at zero bias obtained with the normal tip as a function of the coupling for the first maximum of the wave function for various temperatures (colored points) and for the second maximum at low temperature (crosses).  At given temperature, $dI/dV$ exponentially approaches the quantized value as the $\Gamma$ is increased. The $G_n$ is reached faster at lower temperatures. Inset: Differential conductance at zero bias for the tip connected at the end of the nanowire  and as a function of the temperature for two values of the coupling strength.
At given $\Gamma$, $dI/dV$ exponentially approaches the quantized value as the temperature is decreased. The $G_n$ is reached faster at large $\Gamma$ values. The substrate is in the same configuration as in Fig. \ref{fig:Fig2}. 
}
\label{fig:Fig3}
\end{figure}

\section{Normal metallic STM tip} \label{Normal}

First, we explore how the spatial profile of the differential conductance at zero bias depends on the tunneling rate $\Gamma$ in the case of a normal metallic STM probe, see Fig. \ref{fig:Fig2}.  Importantly, the general feature of the Majorana wave function is clearly captured for all values of $\Gamma$. 
Generally, the stronger $\Gamma$, the larger is the conductance, see Fig. \ref{fig:Fig3}. At lower temperatures, the maximum of conductance comes close to $G_n$ \cite{Anton, Law} if the tip is connected at the end of the wire where the amplitude of the wave function is maximal and, thus, the coupling between the MF and the tip is the strongest.
However, we note that the value $G_n$, predicted for transport via MFs is never reached even if we work with wires that are much longer than the MF localization length such that the MFs do not hybridize \cite{Diego1}. To observe the quantized values, the tunneling should be of the order of the superconducting gap, which is not the regime of a STM spectroscopy experiment, where the tip should not perturb the  system to be measured.  The differential conductance also crucially depends on the temperature $T$. If $\Gamma$ is small compared to $T$, there is a strong effect due to temperature broadening.
By decreasing $T$, we get exponentially close to the value $G_n$ for the same set of parameters, see Fig. \ref{fig:Fig3}. Generally, the smaller the tunneling rate $\Gamma$ is, the higher is the saturation temperature, see the inset in Fig.~\ref{fig:Fig3}.
It is also more difficult to reach $G_n$ if the tip probes one of the next maximums, where the weight of the MF is smaller. Along the same lines, the saturation is achieved faster when the wave function is more localized and each of the maximums has more weight.

For small values of $\Gamma$, the conductance maps could be used to extract the localization lengths by direct fitting \cite{Meyer} (see Appendix \ref{Matching}). In contrast to that, in the strong coupling regime, $G$ saturates at the quantized value $G_n$ and the conductance profile does not replicate the LDOS profile anymore. For example, the decay is no longer purely exponential. However, the main  features are still well captured. For example, in Fig. \ref{fig:Fig2}, there is a characteristic feature at  the site $j=54$  resulting from the interplay of two exponential decays, which can be identified both in the LDOS and in  the differential conductance maps. At the same time, the period of the MF wave function oscillations, determined by the Fermi wavelength, is perfectly captured at all $\Gamma$ values and in  excellent agreement with the period extracted from the LDOS.

\section{Superconducting STM tip}\label{Superconducting}

Next, we look at the differential conductance maps obtained with the superconducting tips. Our formalism is valid only around and above $\Delta_{tip}$ and not in the middle of the gap where the Andreev reflection plays an important role. In what follows, for numerical convenience, we put $\Delta_{tip}=\Delta/10$, however, we checked that using larger values of $\Delta_{tip}$ does not affect the results discussed below.

\begin{figure}[t]
\includegraphics[width=8cm]{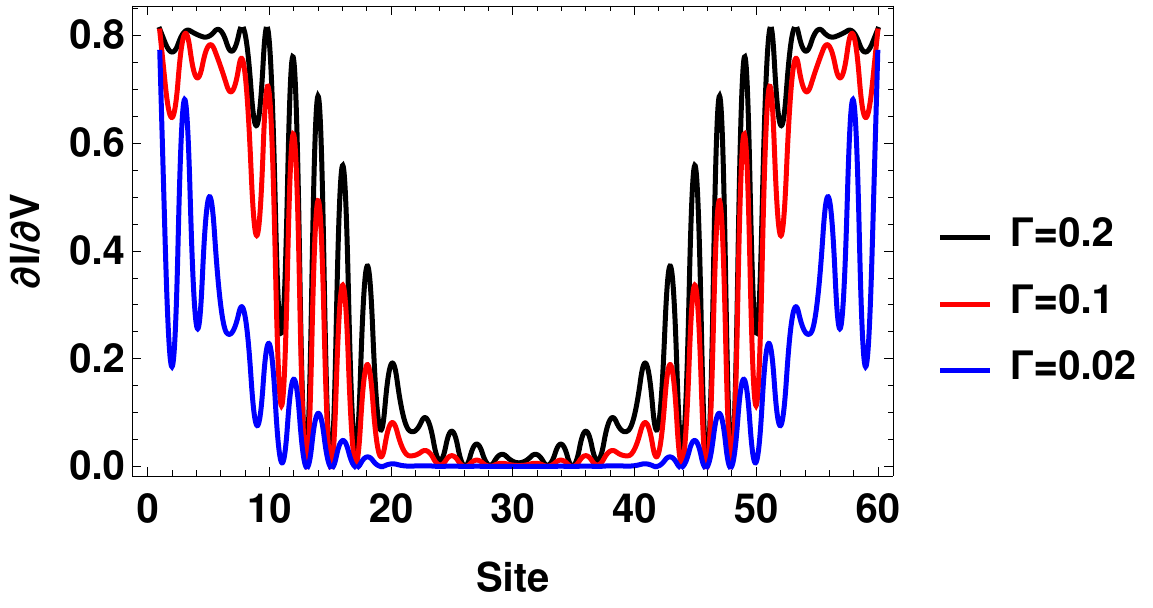}
\caption{The same as in Fig. \ref{fig:Fig2} only for the superconducting tip and at a bias $V_{bias}=\Delta_{tip}$. Again, the differential conductance spatial maps provide information about the MF localization length and the period of oscillations. In contrast to the normal tip data, for large values of $\Gamma$, the signal flattens at the wire ends. }
\label{fig:Fig4}
\end{figure}

\begin{figure}[b]
\includegraphics[width=8cm]{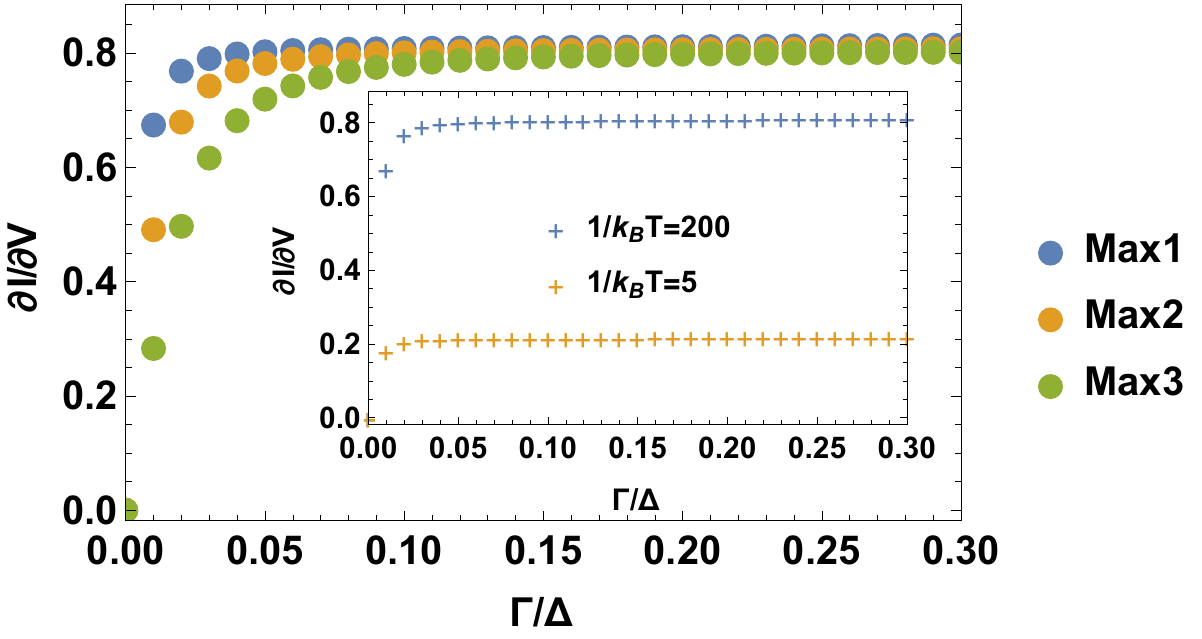}
\caption{The differential conductance as a function of $\Gamma$ for the first three maximums obtained at $V_{bias}=\Delta_{tip}$ for $1/k_B T=200$.  In contrast to the normal tip, the quantized value is reached for smaller $\Gamma$, compare with Fig. \ref{fig:Fig3}. Again, the saturation level is reached faster at the first maximum.
As shown in the inset, where we plot differential conductance for the tip connected to the first site,  the quantized conductance could be reached only at much lower temperatures. The substrate is in the same configuration as in Fig. \ref{fig:Fig2}.}
\label{fig:Fig5}
\end{figure}

\begin{figure}[t]
\includegraphics[width=8cm]{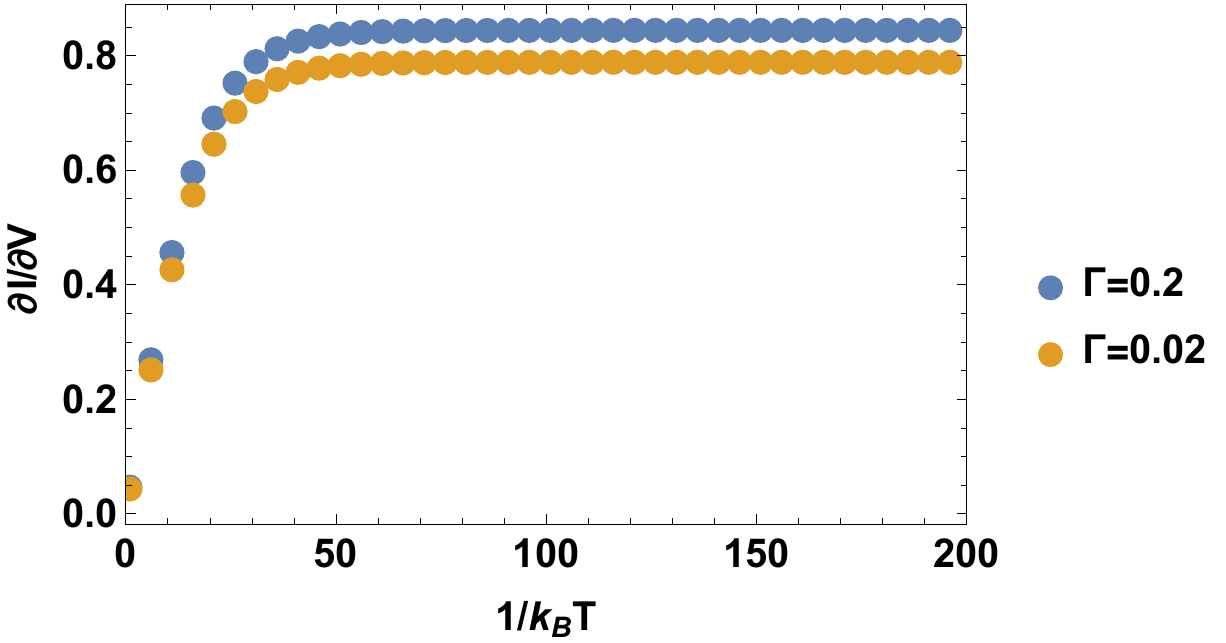}
\caption{Differential conductance for the superconducting tip probing the first site at $V_{bias}=\Delta_{tip}$ as a function of temperature $T$.  The smaller $\Gamma$ is, the smaller is the achieved saturation value. All parameters are the same as in Fig. \ref{fig:Fig4}.}
\label{fig:Fig6}
\end{figure}

Similarly to the normal tip, the superconducting tip measurements (at $V_{bias}=\Delta_{tip}$) give access to the localization length of MFs and the period of their oscillations in the weak tunneling regime (small $\Gamma$), see Fig. \ref{fig:Fig4}.
However, if $\Gamma$ is increased the amplitudes of oscillations tend to be smoothened and the maximum of conductance $G_{sc}=(4-\pi)G_n$ is about to be reached  \cite{Glazman}. In other words, a spatial profile of the differential conductance, corresponding to the MF wave function, first stays almost constant  and then drops abruptly. We note that even if the tip is connected to the first site, the conductance is only slightly smaller than $G_{sc}$, for the same reasons as was discussed for the normal lead ({\it i.e.} finite-size effects, finite temperature, and large tunneling rate).  Generally, the differential conductance is still position-dependent with the largest signal detected at the MF wave function maximums, see Fig. \ref{fig:Fig5}. Importantly, the quantized value for the superconducting tip is reached at much smaller values of $\Gamma$ as it was the case for normal tips.
For example, if the tip is placed above the first maximum, one needs a coupling strength of the order of $2-3\%$ of the gap size instead of the $20\%$ of the normal metal tip. As the tunneling strength is increased further, the differential conductance maps start to develop the plateau in the signal at the wire end, so it gets more difficult to read out the oscillation period, see Fig. \ref{fig:Fig4}. 
Obviously, the competition between the period of the oscillations and the tip resolution plays an important role here. Indeed, we would still observe the oscillations if their period is larger than the tip size (one atom in our case). For example, a differential conductance  could approach $G_{sc}$  for all the maximums but still be zero in-between, which was never the case for a normal probe, see the Appendix \ref{Period_effect}. The MF localization length and period of oscillations depend on system parameters and can be tuned by changing, for example, magnetic field or chemical potential.
We also note that the quantized values $G_{sc}$ is reached only at low temperatures, see the inset of Fig. \ref{fig:Fig5}.
At higher temperature, the conductance also saturates at some value which depends on $T$ as shown in Fig. \ref{fig:Fig6}. Importantly, we find that the differential conductance always gets exponentially close to this value and this behaviour is independent of the transmission coefficient. It can be explained by the fact that the temperature broadening is suppressed in a superconducting STM tip in stark contrast to the normal STM tip, where the signal was always strongly affected by temperature effects. Here this is no longer the case, a quantized peak develops even in a weak tunneling regime.

\begin{figure}[t]
\includegraphics[width=8cm]{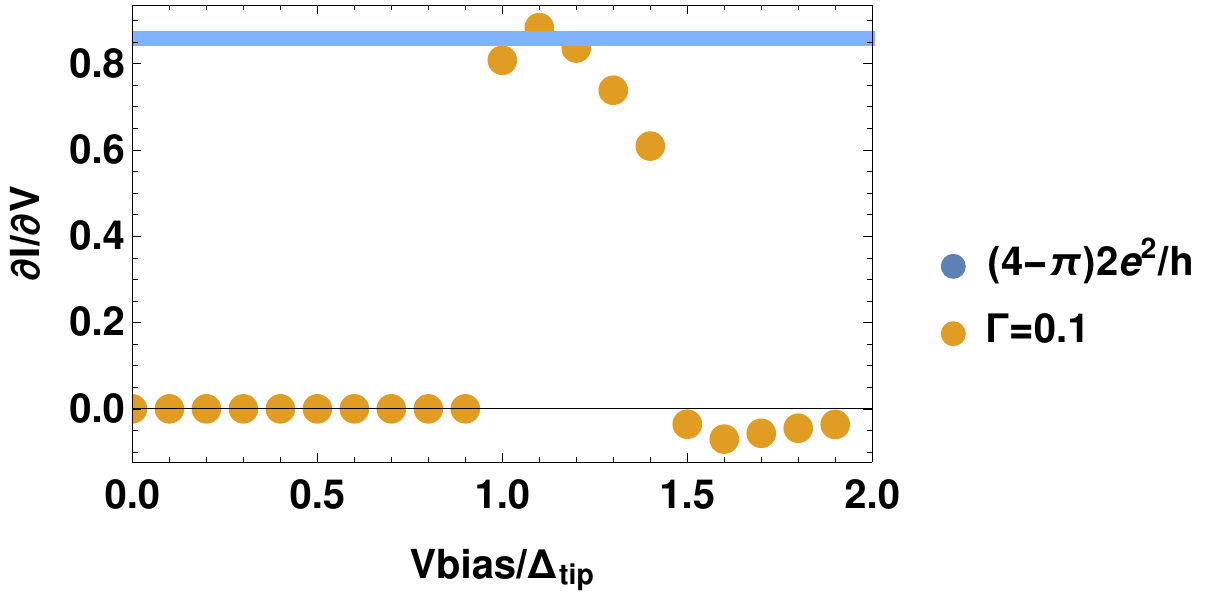}
\caption{The same as in Fig. \ref{fig:Fig4} for the superconducting tip probing at the end of the nanowire. The maximum of the differential conductance at the bias equal to $\Delta_{tip}$ is close to $G_{sc}$ \cite{Glazman, LevyYeyati2}.}
\label{fig:Fig7}
\end{figure}

In addition, unlike the detection with a normal tip where the MF is detected symmetrically around zero bias with a certain width depending on parameters of the system such as the tunneling or disorder for instance, the detection with a superconducting tip occurs at $V_{bias}=\Delta_{tip}$ as mentioned previously, but the shape of the peaks is no longer symmetric. As we can see in Fig. \ref{fig:Fig7}, if the voltage is below $\Delta_{tip}$ there is no current. When the bias reaches $\Delta_{tip}$, a current starts to flow between the substrate and the tip resulting in a peak. If $V_{bias}$ is increased further, the differential conductance  decreases smoothly until it reaches small negatives value \cite{Glazman}.  For further comparison between the two kinds of detection we refer to the Appendix \ref{3Dplots}. 

\section{Discussions and Conclusions} \label{Discussions}

We note that our model based on Rashba SOI and uniform magnetic field is equivalent to the model without any SOI but with a spatially rotating magnetic field \cite{Bernd} produced either externally by local nanomagnets or intrinsically due to RKKY interaction between localized magnetic moments\cite{Daniel, Dominik,RKKY_Basel,RKKY_Simon,RKKY_Franz,Rotating_field}. Thus, our results can be directly applied to the latter systems, for example, to magnetic atom chains on superconducting surfaces \cite{Yazdani,Franke,Meyer}.
We have checked numerically that using such a model does not change any of the results for the differential conductance discussed above.  In Appendix \ref{Experiments_parameters}, we provide additional simulations of strong and weak SOI regimes. We also note that the STM tips can be also used in a similar way to extract the information about the spatial profile of other than MFs types of bound states, such as fractional fermion states or Andreev bound states. \cite{Taddei, Taddei2,Diego,Bound}

It is important to point out that the LDOS of the substrate is generally affected by the tip. In the case of the normal STM tip and weak tunneling limit, the LDOS is unchanged but the amplitude of the MF wave function gets slightly suppressed as the MF tends to leak out into the normal metal \cite{JK_MF,Pascal}. For large value of the tunneling amplitude, the tip becomes a part of the substrate. As a result, one MF effectively disappears by extending into the tip but the MF at the opposite end remains unchanged. In the case of superconducting probe, the behaviour is essentially the same if the bias between the tip and the substrate is equal or larger than $\Delta_{tip}$. 

\acknowledgments

We acknowledge helpful discussions with Daniel Loss. This work was support by the Swiss NSF and  NCCR QSIT.

\appendix

\section{3D plots as a comparison between density of states and differential conductance}\label{3Dplots}

\begin{figure}[t]
\includegraphics[width=5cm]{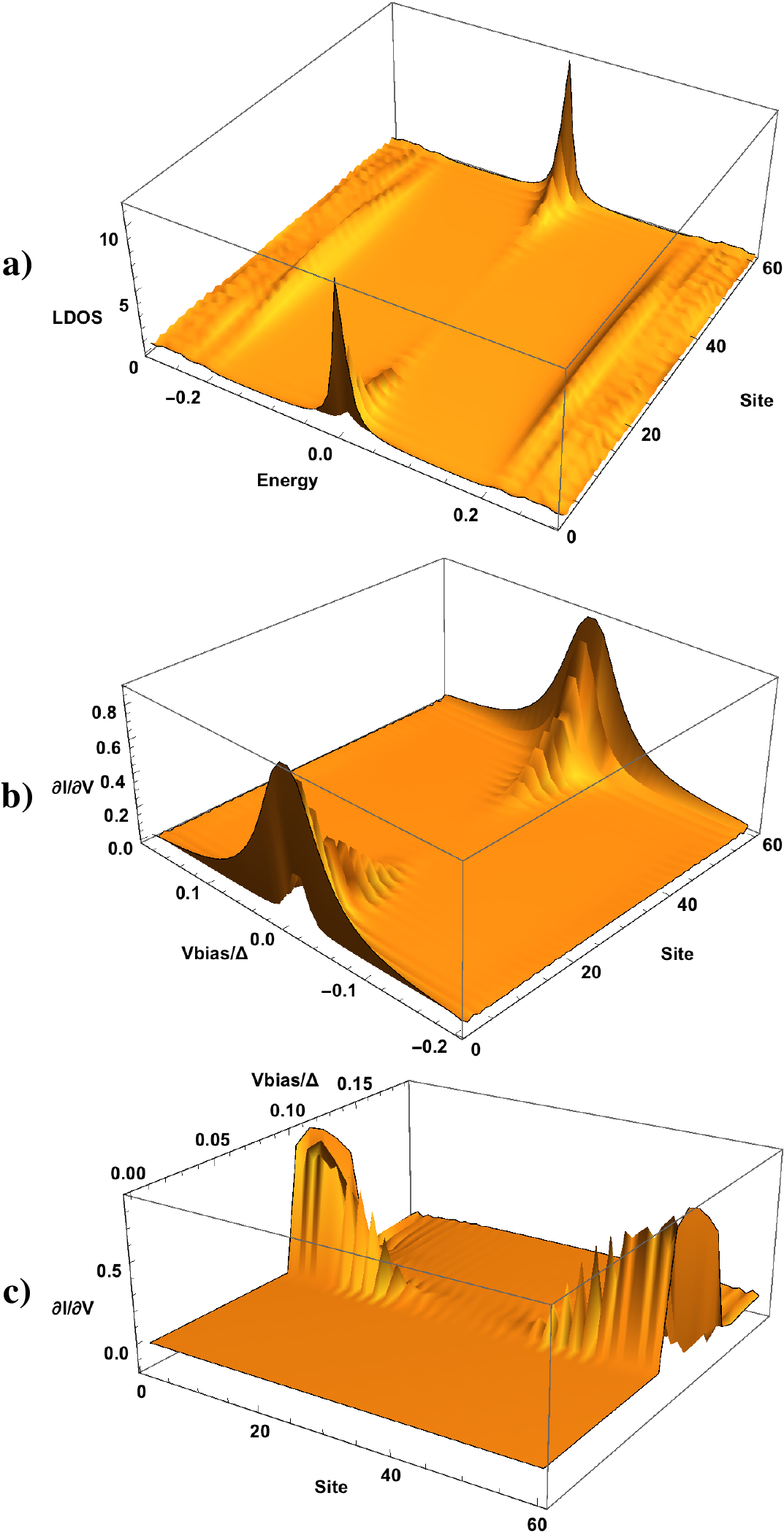}
\caption{ a) LDOS as a function of the position and energy for $\mu=2,V_z=2, \Delta=1, \alpha=0.2$. Two zero-energy modes rise at the end of the nanowire. 
b) Corresponding differential conductance obtained with the normal STM tip as a function of the bias and the position along the nanowire for $\Gamma=0.1$ and $1/k_B T=200$. c) Corresponding differential conductance obtained with the superconducting STM tip as a function of bias voltage and position for $\Gamma=0.1$ and $1/k_B T=200$.}
\label{fig:FigA1}
\end{figure}

In Fig. \ref{fig:FigA1}, we have summarized the main results of our paper where we plot (a) the LDOS as a function of energy and position as well as the differential conductance as a function of the bias voltage and the position obtained with (b) the normal  STM tip and (c) the superconducting STM tip. Both measurements can be used to reconstruct the MF wave function. The important difference concerns the heights of the peak in the differential conductance, which are generally more pronounced in the case of the superconducting case for several reasons, as was discussed in the main text.

\section{Green's function method to calculate the currrent and the conductance}\label{Model_derivation}

In this Appendix, we calculate the current flowing between the tip and the substrate using the Keldysh formalism. To do so, we need to write down all components of the total Green function of the system and the self energy of the tip in the Keldysh space. In order to obtain the Keldysh components of the self-energy, we need to know all the components of the tip Green function in this space. The retarded Green function of the tip is already well known in the literature \cite{LevyYeyati}. Then, it is pretty straightforward to get the advanced one $\tilde{g}_A(\omega)$ and the Keldysh one $\tilde{g}_K(\omega)$ using the following expressions 
\begin{align}
&\tilde{g}_A(\omega)=(\tilde{g}_R(\omega))^\dagger\label{tip_green_1}\\
&\tilde{g}_K(\omega)=(1-2f_\omega)(\tilde{g}_R(\omega)-\tilde{g}_A(\omega))\label{tip_green_2}
\end{align} 
where $f_\omega$ is the standard Fermi-Dirac distribution function (Note that the temperature dependence enters only here). Using the expression of the self energy of the main text combined with Eqs. (\ref{tip_green_1}) and (\ref{tip_green_2}), we are able to get all the components of the self-energy tip in the Keldysh space in the case of normal metal
\begin{align}
&\tilde{\Sigma}^{N}_{R/A}(\omega)=\pm i\Gamma_j \mathds{1}\otimes\begin{pmatrix}1 & 0&\\
0& 1&\\
\end{pmatrix}\\
&\tilde{\Sigma}^{N}_K(\omega)=-2i \Gamma_j \mathds{1}\otimes \begin{pmatrix}
\textrm{tanh}(\beta(\omega-V)/2)&0 \\
0&\textrm{tanh}(\beta(\omega+V)/2)\\
\end{pmatrix} 
\end{align}
where $\mathds{1}$ is the unity matrix in the spin space. The case of superconducting tip is a bit more tricky because of the off-diagonal terms. However, we are not interested in the Andreev reflection processes which can occur inside the gap of the tip but more specifically on what happens close to the gap when $V\approx\Delta_{tip}$. We can thus set these off-diagonal terms corresponding to the Andreev processes to zero \cite{Glazman}. Hence, we can write down the Keldysh components of the superconducting self energy
 \begin{align}
&\tilde{\Sigma}^{S}_{R/A}(\omega)=\Gamma_j \mathds{1}\otimes\begin{pmatrix}\tilde{X}_{R/A}(\omega-V) & 0&\\
0& \tilde{X}_{R/A}(\omega+V)&\\
\end{pmatrix}\\
&\tilde{\Sigma}^{S}_K(\omega)=\Gamma_j \mathds{1}\otimes \begin{pmatrix}
\tilde{X}_{K}(\omega-V)&0 \\
0&\tilde{X}_{K}(\omega+V)\\
\end{pmatrix} 
\end{align} 
with
\begin{align}
&\tilde{X}^{S}_{R/A}(\omega)=-\frac{\Theta(\Delta-|\omega|)\omega}{\sqrt{\Delta^2-\omega^2}}\pm i\frac{\Theta(|\omega|-\Delta)|\omega|}{\sqrt{\omega^2-\Delta^2}}\\
&\tilde{X}^{S}_{K}(\omega)=-2i\frac{\Theta(|\omega|-\Delta)|\omega|}{\sqrt{\omega^2-\Delta^2}}\textrm{tanh}(\beta\omega/2)
\end{align}

\begin{figure}[t]
\includegraphics[width=8cm]{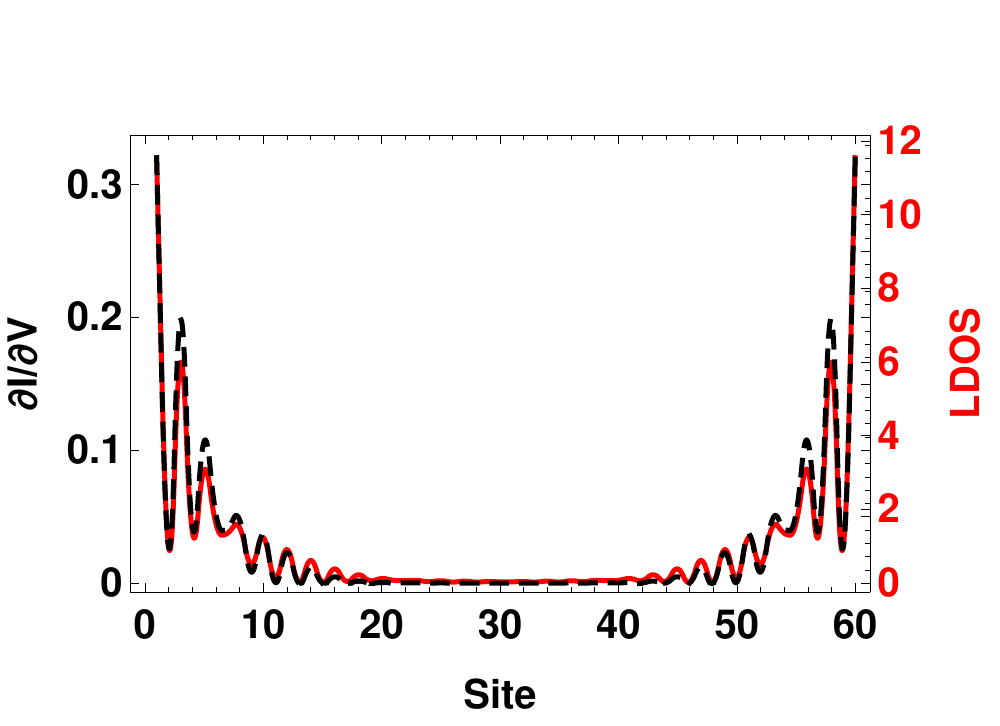}
\caption{ LDOS (red line) and corresponding differential conductance at zero bias (black line) obtained using a normal metal STM tip as a function of position for $\Gamma=0.05$ and $1/k_B T=200$. The substrate is in the following configuration: $\mu=2$, $V_z=2$, $\Delta=1$, and $\alpha=0.2$.}
\label{fig:FigA2}
\end{figure}

Finally, we can get the components of the total Green function using the following expressions
\begin{align}
&\tilde{G}^{-1}_{R/A}(\omega)= \tilde{G}^{-1}_{0R/0A}(\omega)-\tilde{\Sigma}_{R/A}(\omega)\\
&\tilde{G}_{K}(\omega)=\tilde{G}_{0K}(\omega)+\tilde{G}_{R}(\omega)\tilde{\Sigma}_{K}(\omega)\tilde{G}_{A}(\omega)
\end{align}
where $\tilde{G}_{0K}(\omega)=0$ because of the properties of the rotated Keldysh basis. Now we have all the components of the total Green function and the self energy of the tip in the Keldysh space, we can calculate the current flowing from the tip into the substrate. The density of charge reads 
\begin{equation}
\frac{\partial \rho}{\partial t}=\frac{1}{i}[\rho,\tilde{H}], 
\end{equation}
and thus the current can be written as
\begin{equation}
I=-\frac{1}{2}\frac{\partial \rho}{\partial t}.
\end{equation}
Due to the fact that we use the extended Nambu spinor, we add an one half in front of the current density in order to count only once each contributions. 
So, in our case, the current from the tip into the lead is equal to
\begin{equation}
I(t)=\frac{i}{2}\left[\sum_{k}\tilde{\Psi}^\dagger_{k}\tau_z\tilde{\Psi}_{k},\tilde{H}_T(t)\right]=\frac{i}{2}\sum_{k}\tilde{\Psi}^\dagger_{k}e^{i k j}\tau_z\tilde{t_j}(t)\tilde{\psi}_{j}.
\end{equation}
By calculating the partition function of the total system, adding a counting field and taking derivative of it, we are able to extract the current in term of the total Green function of the system and the self energy of the tip (This model is well known and has been used in \cite{Zazunov,Chevallier} for instance). The final answer for the current reads
\begin{align}\label{current_time}
&\langle I(t)\rangle=\frac{1}{2}\textrm{Tr}\{\tau_z \\
&\times\int^{+\infty}_{-\infty} dt'[\tilde{G}_R (t,t')\tilde{\Sigma}_K(t',t)+\tilde{G}_K(t,t')\tilde{\Sigma}_A(t',t)]\}.\notag
\end{align}
In the stationary regime, we can write down immediately the dc current flowing between the tip and the substrate by taking the Fourier transform of Eq. (\ref{current_time}) and returning to the physical units
\begin{equation}
I_{dc}=\frac{e}{2\hbar}\textrm{Tr}\{\tau_z \int^{+\infty}_{-\infty} \frac{d\omega}{2\pi} [\tilde{G}_R (\omega)\tilde{\Sigma}_K(\omega)+\tilde{G}_K(\omega)\tilde{\Sigma}_A(\omega)]\}.
\end{equation}
The differential conductance reads
\begin{equation}
G(V)=\frac{\partial I_{dc}}{\partial V}.
\end{equation}

\begin{figure}[t]
\includegraphics[width=8cm]{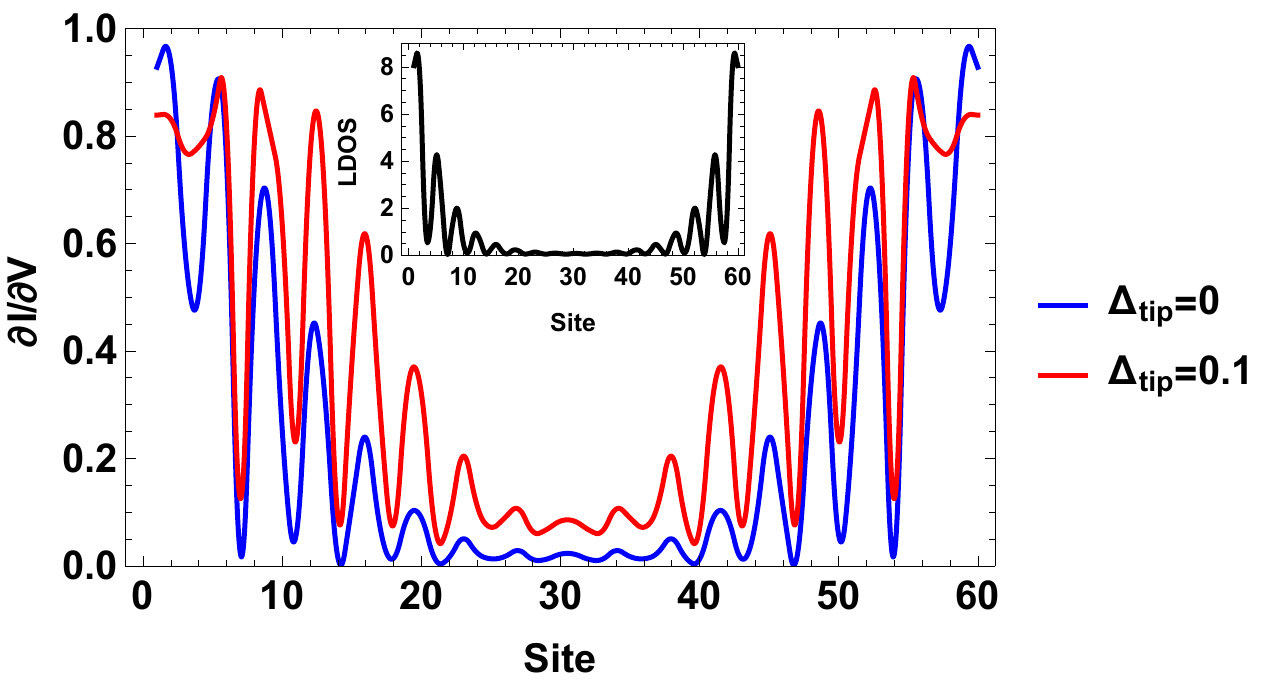}
\caption{Differential conductance at zero bias ($V_{bias}=\Delta_{tip}$) obtained with a normal metal (superconducting) STM tip as a function of position in the strong coupling regime $\Gamma=0.2$ and $1/k_B T=200$. The substrate is in the following configuration: $\mu=1/2$, $V_z=2$, $\Delta=1$, and $\alpha=0.2$. The insert shows the LDOS.}
\label{fig:FigA3}
\end{figure} 

\section{Matching the LDOS and the differential conductance in tunneling limit} \label{Matching}

In Fig. \ref{fig:FigA2}, we have plotted the LDOS and the corresponding differential conductance in the tunneling limit ($\Gamma=0.05$). For such values of $\Gamma$, the conductance maps exactly the MF wavefunction density profile  and could be used to extract the localization lengths. As noted in the main text, this is not the case in the strong tunneling limit.

\section{Effect of the period of the oscillations on the differential conductance} \label{Period_effect}

To manipulate the period of the oscillations, we can change the Fermi wave vector $k_F$ by tuning $\mu$ (another way would be to tune the magnetic field for instance). In Fig. \ref{fig:FigA3}, we have plotted the LDOS and the differential conductance obtained with normal metal and superconducting probes for the chemical potential $\mu=1/2$. 

\begin{figure}[t]
\includegraphics[width=8cm]{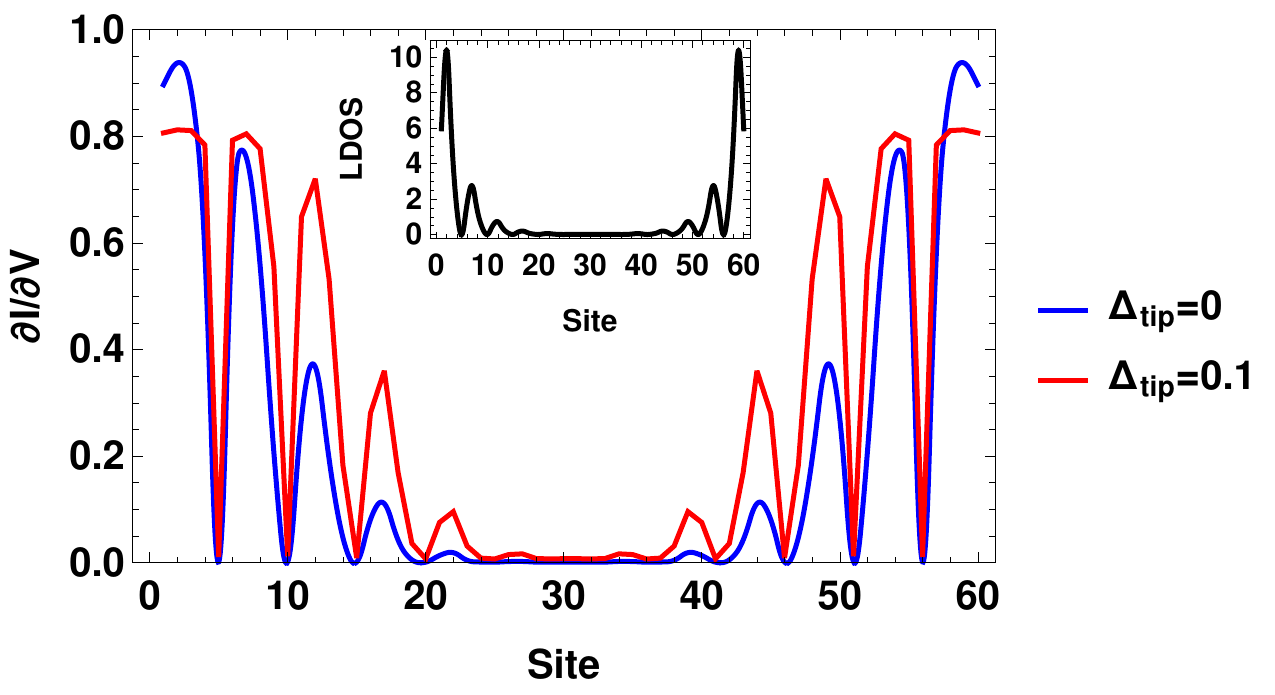}
\caption{Differential conductance at $V_{bias}=0$ ($V_{bias}=\Delta_{tip}$) obtained with a normal metal (superconducting) STM tip as a function of position in the strong coupling regime $\Gamma=0.2$ and $1/k_B T=200$. The substrate is in the strong SOI configuration: $t=5$, $\mu=10$, $V_z=2$, $\Delta=1$, and $\alpha=3.5$. The insert shows the LDOS.}
\label{fig:FigA4}
\end{figure}

The differential conductance in the case of a normal metal probe, as expected, is in good agreement with the LDOS. Generally, the differential conductance catches perfectly the oscillation period. In the case of a superconducting tip, the saturation plateau is absent, as the signal periodically drops almost to zero. What is important is the size of the tip compared to the oscillations period. Experimentally, MF wave functions have a large period of oscillations because the SOI in InAs or InSb is believed to be weak, thus, the saturation level should not be achieved. 
Alternatively, one can use a nanowire without SOI but in presence of rotating magnetic field. In that case it is possible to tune the strength and the period of oscillations in order to have such a substrate with large localization length and large oscillation period, corresponding to the distance between magnets \cite{Bernd,Fle,graphene,MoS2}.

\begin{figure}[b]
\includegraphics[width=8cm]{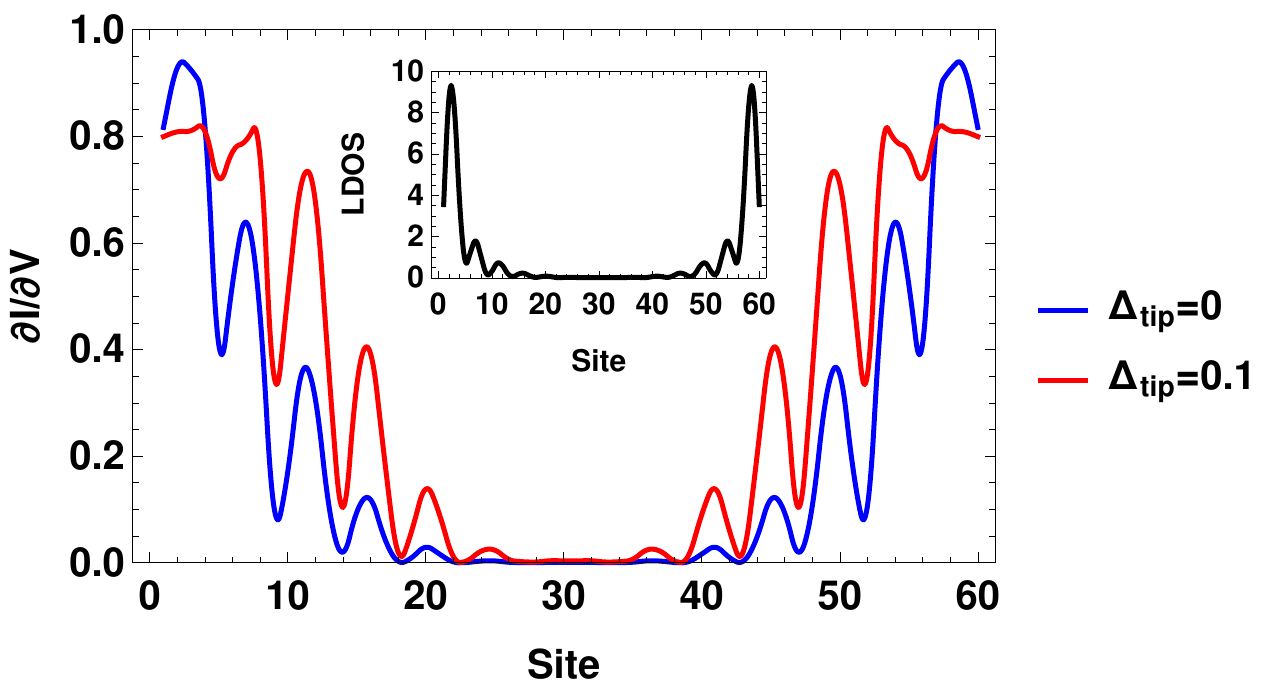}
\caption{Differential conductance at $V_{bias}=0$ ($V_{bias}=\Delta_{tip}$)  obtained with a normal metal (superconducting) STM tip as a function of position in the strong coupling regime $\Gamma=0.2$ and $1/k_B T=200$.
The substrate is in the weak SOI configuration: $t=5$, $\mu=10$, $V_z=2$, $\Delta=1$, and $\alpha=1.2$. The insert shows the LDOS.}
\label{fig:FigA5}
\end{figure} 

\section{Strong and Weak spin-orbit regime configurations} \label{Experiments_parameters}

In this section, we perform the simulations for the parameters in the weak ($\alpha=1.2$, $E_{SO}=\alpha^2/t\approx0.3\Delta$) and strong ($\alpha=3.5$, $E_{SO}=\alpha^2/t\approx2.5\Delta$) SOI regimes with ratios between key parameters close to the experimental settings: $t=5$, $\mu=10$, $V_z=2$, $\Delta=1$. In both configurations, $\Delta/t=0.2$ and the temperature is $1/k_B T\approx 200$ (in units of $\Delta$). From results represented in Figs. \ref{fig:FigA4} and \ref{fig:FigA5},  we conclude that the presence of spatial oscillations, temperature dependence, and saturation of conductance is a general feature of MF nanowires and can be observed with both metallic and superconducting STM tips.

\end{document}